\documentclass[twocolumn]{IEEEtran}
\usepackage{graphicx}
\usepackage{amsmath,amssymb,amsthm}

\usepackage{tikz}
\usetikzlibrary{arrows}
\usepackage{pgfplots}

\newcommand{\av}{\ensuremath{\mathbf{a}}}
\newcommand{\Gm}{\ensuremath{\mathbf{G}}}
\newcommand{\Jm}{\ensuremath{\mathbf{J}}}

\newcommand{\nv}{\ensuremath{\mathbf{n}}}

\newcommand{\uv}{\ensuremath{\mathbf{u}}}

\newcommand{\vv}{\ensuremath{\mathbf{v}}}
\newcommand{\wv}{\ensuremath{\mathbf{w}}}

\graphicspath{{Figures/}}

\title{
Localization in Autonomous Vehicles \\
Using a Generalized Inner Product}

\author{\IEEEauthorblockN{Samuel Todd Flanagan,
Drupad K. Khublani, \\
Jean-Francois Chamberland,
Siddharth Agarwal,
Ankit Vora}
\thanks{
This material is based, in part, upon work supported by Ford Autonomous Vehicles LLC.
S. T. Flanagan, D. K. Khublani, and J.-F. Chamberland are with the Department of Electrical and Computer Engineering, Texas A{\&}M University, College Station, TX 77843, USA (emails: \{stflanagan, dkhublani, chmbrlnd\}@tamu.edu).
S. Agarwal and A. Vora are with Ford Autonomous Vehicles LLC, Dearborn, MI 48126, USA (emails: \{sagarw20, avora3\}@ford.com).
}}

\pgfplotsset{compat=1.14} % overleaf said to put this here, changed to 1.14 for arXiv

\begin{document}

\maketitle

\begin{abstract}
Fine localization in autonomous driving platforms is a task of broad interest, receiving much attention in recent years.
Some localization algorithms use the Euclidean distance as a similarity measure between the local image acquired by a camera and a global map, which acts as side information.
The global map is typically expressed in terms of the coordinate system of the road plane.
Yet, a road image captured by a camera is subject to distortion in that nearby features on the road have much larger footprints on the focal plane of the camera compared with those of equally-sized features that lie farther ahead of the vehicle.
Using commodity computational tools, it is straightforward to execute a transformation and, thereby, bring the distorted image into the frame of reference of the global map.
However, this non-linear transformation results in unequal noise amplification.
The noise profile induced by this transformation should be accounted for when trying to match an acquired image to a global map, with more reliable regions being given more weight in the process.
This physical reality presents an algorithmic opportunity to improve existing localization algorithms, especially in harsh conditions.
This article reviews the physics of road feature acquisition through a camera, and it proposes an improved matching method rooted in statistical analysis.
Findings are supported by numerical simulations.
\end{abstract}

\section{Introduction}

Autonomous vehicles have received considerable attention in recent years, with several companies deploying prototypes on public roadways.
A common task for various autonomous platforms is fine localization at the scale of centimeters.
Localization is the process by which an autonomous vehicle uses sensor data to determine its position within a map.
Mathematically, this can be accomplished by calculating the inner product of sensor data and candidate sections of the map.
The candidate section that produces the maximum or minimum inner product, depending on the formulation, is the most likely match.
Inner products have been used in the past for matching in both localization and image registration, e.g.,~\cite{kruger1998image,konolige1999markov}.

Much of the existing results in this area focus on simultaneous localization and mapping (SLAM).
Many implementations of SLAM have been developed over the past two decades.
In 2001, Dissanayake et al.\ proposed a solution to the SLAM problem using an estimation-theoretic or Kalman filter based approach~\cite{dissanayake2001solution}.
Since then, implementations such as DP-SLAM~\cite{eliazar2003dp}, Atlas~\cite{bosse2004simultaneous}, and Graph SLAM~\cite{thrun2006graph} have been proposed.
A current industry standard for SLAM, described in \cite{levinson2007map}, is a version of Graph SLAM that achieves ``reliable real-time localization with accuracy in the 10-cm range.''
In a subsequent article, the authors propose improvements such as using probabilistic maps to represent the environment that ``increased robustness to environmental changes and dynamic obstacles,'' while retaining similar accuracy \cite{levinson2010robust}.

These autonomous vehicle systems use high-quality sensor arrays in favorable conditions to provide near-noiseless data for localization.
However, as autonomous vehicles move closer to production, lower quality sensors will likely be utilized to reduce cost.
Inexpensive cameras have been used in SLAM research (visual SLAM), yet proposed solutions struggle in challenging conditions~\cite{fuentes2015visual}.
In this paper, we use signal processing techniques to develop a novel algorithm that leverages a generalized inner product for localization.
The proposed scheme outperforms the standard inner product for matching, with significant gains under adverse conditions.

\section{Problem Formulation}

This work focuses on localization using a single camera with a predetermined global map that accurately represents a top-down view of the environment.
Local images are captured in real-time, transformed, scaled, and matched with a section of the map.
Herein, we treat gray-scale images, yet the proposed techniques can be extended to color images, image gradients, and processed images with minor changes to the implementation.

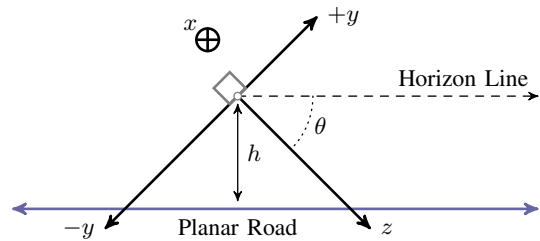
\begin{figure}[b]
\centerline{\begin{tikzpicture}[
  font=\small, >=stealth', line width=1.0pt, line cap=round
]

% planar road
\draw[<->, blue!40!black!60] (-3, -1.5) -- (0, -1.5) node[anchor=north, text=black] {Planar Road} -- (4, -1.5);

% height line
\draw[<->, line width=0.5] (0, -0.1) -- (0, -0.75) node[anchor=west, text=black] {$h$} -- (0, -1.4);

% theta arc
\draw[dotted, line width=0.5] (1, 0) arc (0:-45:1);
\draw (1.1, -0.4) node[text=black] {$\theta$};

% physical coordinate system
% x axis
\draw (-0.4, 0.75) circle (0.15) node[anchor=south east] {$x$};
\draw (-0.55, 0.75) -- (-0.25, 0.75);
\draw (-0.4, 0.6) -- (-0.4, 0.9);
% y axes
\draw[->] (0, 0) -- (45:1.5) node[anchor=west] {$+y$};
\draw[->] (0, 0) -- (-135:2.5) node[anchor=east] {$-y$};
% z axis
\draw[->] (0, 0) -- (-45:2.5) node[anchor=west] {$z$};

% horizon line
\draw[dashed, ->, line width=0.5] (0, 0) -- (3, 0) node[anchor=south, text=black] {Horizon Line} -- (4, 0);

% pinhole camera
\draw[gray, rotate around={-45:(0, 0)}] (-0.3, -0.15) rectangle (0, 0.15);
% \draw (0, 1) node[anchor=south] {Focal Plane};
\draw[gray, fill=white, line width=0.5] (0, 0) circle (0.05);

% camera label
% \draw[->, line width=0.5] (-2, -0.5) node[anchor=north east] {Pinhole Camera} -- (-0.4, 0);

\end{tikzpicture}}
\caption{This diagram illustrates the physical coordinate system and the camera orientation.}
\label{figure:PhysicalCoordinateSystem}
\end{figure}

Our model has a focal plane and physical coordinate system with 2 and 3 dimensions, respectively.
We use $x$, $y$, and $z$ for the physical coordinate system and $\tilde{x}$ and $\tilde{y}$ for the focal plane of the camera.
The origin of the physical coordinate system is centered on the pinhole of the camera, height $h$ above a planar road as shown in Fig.~\ref{figure:PhysicalCoordinateSystem}.
Parameter $\theta$ denotes the angle between the $z$-axis and the horizon.

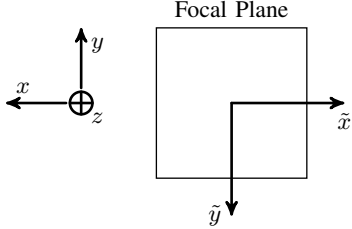
\begin{figure}[tbh]
\centerline{\begin{tikzpicture}[
  font=\small, >=stealth', line width=1.0pt, line cap=round
]

% focal plane
\draw[line width=0.5] (-1, -1) rectangle (1, 1);
\draw (0, 1) node[anchor=south] {Focal Plane};

% focal plane coordinate system
\draw[->] (0, 0) -- (0, -1.5) node[anchor=east] {$\tilde{y}$};
\draw[->] (0, 0) -- (1.5, 0) node[anchor=north] {$\tilde{x}$};

% physical coordinate system
% z axis
\draw (-2, 0) circle (0.15) node[anchor=north west] {$z$};
\draw (-1.85, 0) -- (-2.15, 0);
\draw (-2, -0.15) -- (-2, 0.15);
% x and y axes
\draw[->] (-2, 0.2) -- (-2, 1) node[anchor=north west] {$y$};
\draw[->] (-2.2, 0) -- (-3, 0) node[anchor=south west] {$x$};

\end{tikzpicture}}
\caption{The physical coordinate system is selected to align, partly, with the coordinate system of the focal plane, as shown above.}
\label{figure:FocalPlaneCoordinateSystem}
\end{figure}

In our model, we use a pinhole camera with focal length, $f$, the distance between the pinhole and the focal plane.
It is worth noting that, for this initial treatment, we disregard barrel and pincushion distortion in our analysis.
The origin of the focal plane coordinate system is located at the center of the focal plane, $(0, 0, -f)$ in physical coordinates.
Figure~\ref{figure:FocalPlaneCoordinateSystem} shows the orientation of the coordinate systems.
$\tilde{x}$ and $\tilde{y}$ are antiparallel to $x$ and $y$ which allows us to effectively ignore image inversion in the following sections.

\begin{figure}[thpb]
\centering
\includegraphics[scale=0.2]{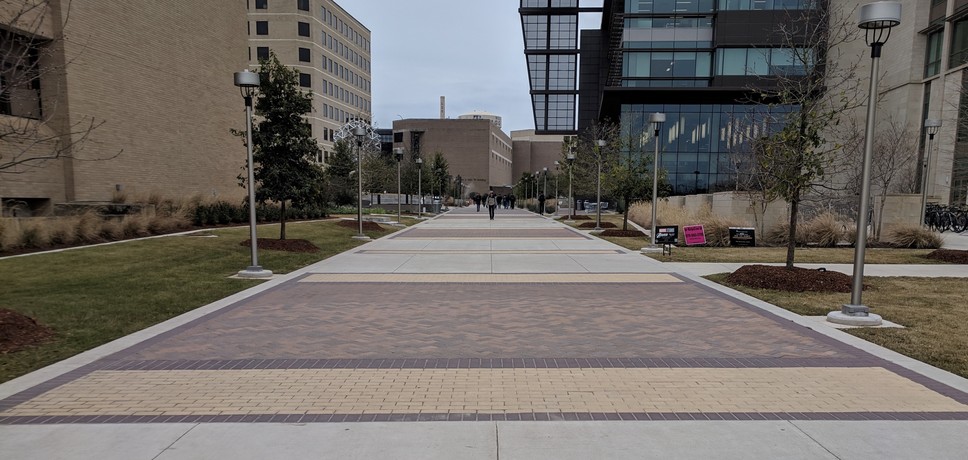}
\caption{This image captures the effects of a camera perspective.}
\label{figure:ExamplePerspective}
\end{figure}

A perspective transformation models how objects appear when viewed from different positions~\cite{hearn2011computer}.
Figure~\ref{figure:ExamplePerspective} highlights several aspects of the perspective transformation at hand.
Nearby objects appear larger than those farther away.
Lines parallel to the view plane remain parallel.
Lines not parallel to the view plane are distorted.

For our model, the perspective transformation that describes how objects are projected onto the focal plane, is given by
\begin{xalignat}{2} \label{equation:Tilde}
\tilde{x} &= {fx}/{z}  &
\tilde{y} &= {fy}/{z} .
\end{xalignat}

\section{Focal Plane Geometry}
\label{section:FocalPlaneGeometry}

In this section, we develop a relationship between the area of a section of road and the footprint of its image on the focal plane.
We describe the road surface shown in Fig.~\ref{figure:PhysicalCoordinateSystem} with free variables $x$ and $z$.
Consequently, $y$ can be expressed as $z \tan \theta - h \sec \theta$,
where $\theta$ is the angle between the $z$ axis and the horizon.
We use this relation to rewrite part of \eqref{equation:Tilde} as
\begin{equation} \label{equation:yTildeFunctionOfZ}
\tilde{y} = f \tan \theta - ({f h}/{z}) \sec \theta .
\end{equation}
We also need to express $z$ as a function of $\tilde{y}$ for our discussion in Section~\ref{section:ImageAcquisitionNoise}.
This relation is readily found to be
\begin{equation} \label{equation:z}
z = \frac{f h}{f \sin \theta - \tilde{y} \cos \theta} .
\end{equation}

\begin{figure}[tbh]
\begin{tikzpicture}[
  font=\small, >=stealth', line width=1.0pt, line cap=round
]

\foreach \z in {1,2,3,4,5} {
  \foreach \x in {-2,-1,0,1,2} {
    \draw (\x, 1) -- (\x, 5);
    \draw[dashed] (\x, 5) -- (\x, 5.5);
    \draw (-2, \z) -- (2, \z);
  }
}
\draw[fill=lightgray] (1,1) rectangle (2,2);
\draw[fill=gray] (-1,2) rectangle (0,3);
\draw[fill=darkgray] (0,3) rectangle (1,4);
\draw[fill=black] (-2,4) rectangle (-1,5);

\end{tikzpicture} \hfill \begin{tikzpicture}[
  font=\small, >=stealth', line width=1.0pt, line cap=round
]

\def\mytheta{-20}

\foreach \z in {1,2,3,4,5,6,7,8} {
  \draw ({-2/\z}, {tan(\mytheta) - (2*sec(\mytheta)/\z)}) -- ({2/\z}, {tan(\mytheta) - (2 * sec(\mytheta)/\z)});
}
\foreach \x in {-2,-1,0,1,2} {
  \draw (\x, {tan(\mytheta) - 2*sec(\mytheta)}) -- ({\x/8}, {tan(\mytheta) - (2 * sec(\mytheta)/8)});
  \draw ({\x/8}, {tan(\mytheta) - (2 * sec(\mytheta)/8)}) -- ({\x/15}, {tan(\mytheta) - (2 * sec(\mytheta)/15)});
}

\draw[fill=black] (-2/4,{tan(\mytheta) - (2 * sec(\mytheta)/4)})
-- (-1/4,{tan(\mytheta) - (2 * sec(\mytheta)/4)})
-- (-1/5,{tan(\mytheta) - (2 * sec(\mytheta)/5)})
-- (-2/5,{tan(\mytheta) - (2 * sec(\mytheta)/5)})
-- (-2/4,{tan(\mytheta) - (2 * sec(\mytheta)/4)});

\draw[fill=darkgray] (1/3,{tan(\mytheta) - (2 * sec(\mytheta)/3)})
-- (1/4,{tan(\mytheta) - (2 * sec(\mytheta)/4)})
-- (0,{tan(\mytheta) - (2 * sec(\mytheta)/4)})
|- (1/3,{tan(\mytheta) - (2 * sec(\mytheta)/3});

\draw[fill=gray] (-1/2,{tan(\mytheta) - (2 * sec(\mytheta)/2)})
-- (-1/3,{tan(\mytheta) - (2 * sec(\mytheta)/3)})
-- (0,{tan(\mytheta) - (2 * sec(\mytheta)/3)})
|- (-1/2,{tan(\mytheta) - (2 * sec(\mytheta)/2)});

\draw[fill=lightgray] (2,{tan(\mytheta) - (2 * sec(\mytheta))})
-- (1,{tan(\mytheta) - (2 * sec(\mytheta)/2)})
-- (1/2,{tan(\mytheta) - (2 * sec(\mytheta)/2)})
-- (1,{tan(\mytheta) - (2 * sec(\mytheta))})
-- (2,{tan(\mytheta) - (2 * sec(\mytheta))});

\end{tikzpicture}
\caption{The diagram on the left showcases an arbitrary road grid placed in front of an autonomous vehicle.
The area of every square is kept constant.
On the right, the drawing illustrates how the same grid pattern appears on the focal plane of the camera.
While all the squares share the same physical area, their footprints on the focal plane differ.}
\label{figure:GridRoadPlane}
\end{figure}
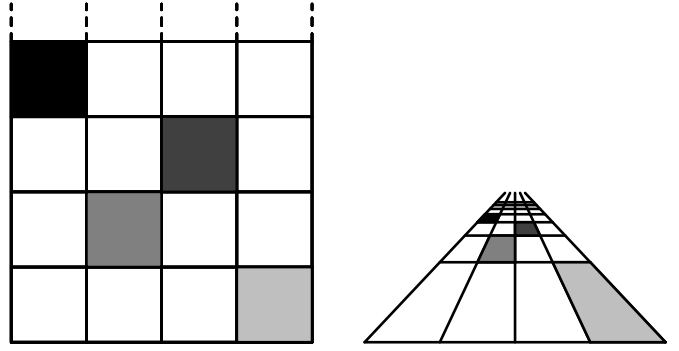

Figure~\ref{figure:GridRoadPlane} depicts an arbitrary grid pattern that lies in front of the autonomous vehicle.
It also shows the same grid pattern as acquired by the focal plane of the camera.
This type of image distortion should be familiar to the reader because an analogous process governs human vision.
It is especially pertinent to note that squares of equal area in the physical world can have vastly different footprints on the focal plane of the camera.
In particular, squares that are farther away correspond to much smaller regions on the focal plane.
This phenomenon has repercussions both in terms of signal acquisition and noise corruption.

In reference to calculus, the absolute value of the Jacobian determinant at point $(x_1,z_1)$ expresses how much the area near $(x_1,z_1)$ expands or contracts when projected onto the focal plane.
The Jacobian of the perspective transformation, governed by \eqref{equation:Tilde} and \eqref{equation:yTildeFunctionOfZ}, is given by
\begin{equation} \label{equation:Jacobian}
\Jm = \begin{bmatrix}
\frac{\partial \tilde{x}}{\partial x} & \frac{\partial \tilde{y}}{\partial x} \\
\frac{\partial \tilde{x}}{\partial z} & \frac{\partial \tilde{y}}{\partial z} \\
\end{bmatrix}
= \begin{bmatrix} \frac{f}{z} & 0 \\
- \frac{f x}{z^2} & - \frac{f h}{z^2} \sec \theta \end{bmatrix} .
\end{equation}
The Jacobian determinant, $\operatorname{det}(\Jm)$, is found to be
\begin{equation} \label{equation:DeterminantJacobian}
\operatorname{det}(\Jm)
= - {f^2 h \sec \theta}/{z^3} .
\end{equation}
Using \eqref{equation:DeterminantJacobian} we compute the relationship between the area of a rectangular region $\mathcal{R} = [x_{\mathrm{l}}, x_{\mathrm{u}}] \times [z_{\mathrm{l}}, z_{\mathrm{u}}]$, which lies on the road in front of the autonomous vehicle, and the area of its projection on the focal plane as
\begin{equation} \label{equation:TildeA}
\tilde{\mathcal{A}} = f^2 h \sec \theta \frac{(x_{\mathrm{u}} - x_{\mathrm{l}})}{2} \left( \frac{1}{z_{\mathrm{l}}^2} - \frac{1}{z_{\mathrm{u}}^2} \right) .
\end{equation}

Having developed this relationship, we can characterize the noise pattern associated with the perspective transformation.

\section{Image Acquisition and Noise}
\label{section:ImageAcquisitionNoise}

The purpose of this section is to describe how the transformation from the physical world to the focal plane of the camera impacts the quality of the image.
Suppose that, at a given time, features on the road surface are captured by $g(x,z)$.
Then, at the same instant, the projection of the road onto the focal plane of the camera can be expressed, using \eqref{equation:Tilde} and \eqref{equation:z}, as
\begin{equation*}
\tilde{g} ( \tilde{x}, \tilde{y} )
= g \left( \frac{h \tilde{x}}{f \sin \theta - \tilde{y} \cos \theta}, \frac{h \tilde{y}}{f \sin \theta - \tilde{y} \cos \theta} \right) .
\end{equation*}
The signal captured on the focal plane of the camera as a function of location $( \tilde{x}, \tilde{y} )$ is given by $\tilde{g}( \tilde{x}, \tilde{y} ) + N ( \tilde{x}, \tilde{y} )$,
where $N$ is two-dimensional white noise with power spectral density $N_0$.
Consequently, the aggregate signal over a region becomes
\begin{equation*}
\begin{split}
&S =
\underbrace{\iint_{\tilde{R}} \tilde{g}( \tilde{x}, \tilde{y} ) d\tilde{x} d\tilde{y}}_{\text{amplitude}}
+ \underbrace{\iint_{\tilde{R}} N ( \tilde{x}, \tilde{y} ) d\tilde{x} d\tilde{y}}_{\text{noise}} .
\end{split}
\end{equation*}
When the CCD sensor is operating within its linear region, as opposed to saturation, thermal noise is well-modeled as an additive zero-mean Gaussian random variable with variance $\sigma^2 = \tilde{\mathcal{A}} N_0$.
The 2D power spectral density parameter $N_0$ implicitly depends on a number of factors, yet it remains constant over the focal plane of the camera.
To gain further intuition into the impact of the perspective transformation over performance, we examine a simplified model where the spatial signal has a constant magnitude in the physical world.

\subsection{Constant Amplitude Regions}
\label{subsection:ConstantMagnitudeSignal}

Consider a situation where a rectangular area on the planar road possesses a constant magnitude.
Assume this rectangle corresponds to region $\mathcal{R} = [x_{\mathrm{l}}, x_{\mathrm{u}}] \times [z_{\mathrm{l}}, z_{\mathrm{u}}]$.
That is, $g(x,z) = a$ for $(x,z) \in \mathcal{R}$.
We wish to compute the effective signal-to-noise ratio (SNR) corresponding to this area as observed by the focal plane.
We begin by computing the energy of the signal component, which is equal to
\begin{equation} \label{equation:FocalPlaneValue}
\begin{split}
\left( \iint_{\tilde{\mathcal{R}}} \tilde{g}( \tilde{x}, \tilde{y} ) d\tilde{x} d\tilde{y} \right)^2
= \left( \iint_{\tilde{\mathcal{R}}} a d\tilde{x} d\tilde{y} \right)^2
= a^2 \tilde{\mathcal{A}}^2 .
\end{split}
\end{equation}
An explicit expression for $\tilde{\mathcal{A}}$ appears in \eqref{equation:TildeA}.
By assumption, the noise component is independent of $\tilde{g}( \tilde{x}, \tilde{y} )$.
As mentioned above, the noise is zero-mean with variance $\tilde{\mathcal{A}} N_0$.
The effective signal-to-noise ratio for the component of the image corresponding to region $\mathcal{R}$ is then given by
\begin{equation} \label{equation:SNR}
\operatorname{SNR}
= \frac{a^2 \tilde{\mathcal{A}}}{N_0}
= \frac{a^2}{N_0} \frac{f^2 h}{\cos \theta} \frac{(x_{\mathrm{u}}-x_{\mathrm{l}})}{2} \left( \frac{1}{z_{\mathrm{l}}^2} - \frac{1}{z_{\mathrm{u}}^2} \right) .
\end{equation}
The main insight is that the SNR decreases dramatically as a function of $z$.
When matching a local image to a global map, this inherent phenomenon should be taken into consideration.
This is especially true for situations where average SNR is low, such as poor light conditions or images acquired by inexpensive cameras.
To further illustrate this point, we explore the structure of an optimal decision rule for tesselated images with piecewise constant amplitude.

\subsection{Maximal Ratio Combining}
\label{subsection:MaximalRatioCombining}

Having developed a suitable characterization of the SNR for a uniform region, we turn to the scenario where the road plane is partitioned into multiple rectangular regions in a manner akin to Fig.~\ref{figure:GridRoadPlane}.
To facilitate analysis, we calibrate observations to be centered around zero.
We assume every square is randomly assigned a constant amplitude value independent of other regions.
Because the road plane is acquired by a pinhole camera, different square regions feature different SNRs, as described above.
The ensuing localization process entails matching the acquired image to a quantized global map.
Under current assumptions, the localization task becomes a canonical vector Gaussian classification problem.

The tesselated road portion acquired by the camera can be reshaped into a vector, which we denote by $\av$.
This gives rise to a noisy observation vector
\begin{equation*}
\vv = \av + \nv ,
\end{equation*}
where $\av$ is the amplitude vector and $\nv$ denotes an additive Gaussian noise vector.
In this equation, the value of indexed element $a_k$ in $\av = (a_1, \ldots, a_m)$ corresponds to the amplitude of a rectangular region on the road surface.
Component $n_k$, which accounts for the effects of the noise process associated with the CCD sensor over the footprint of rectangular region $\tilde{\mathcal{R}}_k$, has mean zero and variance
\begin{equation*}
\sigma_k^2 = {N_0}/{\tilde{\mathcal{A}}_k} .
\end{equation*}
Since the tesselated regions do not overlap on the focal plane, the components of noise vector $\nv$ are independent from one another.
The effective SNR for region~$\mathcal{R}_k$ is then given by
\begin{equation*}
\operatorname{SNR}_k = {a_k^2 \tilde{\mathcal{A}}_k}/{N_0}
\end{equation*}
which is equivalent to \eqref{equation:SNR}.

In estimating the location of the vehicle, we assume that candidate locations are defined at the level of the road tesselation, or a constant integer multiple thereof.
This way, candidate vectors all share a similar structure with regions of constant amplitude.
The likelihood for candidate grid location $\hat{\uv}$ is
\begin{equation*}
\begin{split}
&\mathcal{L} \left( \hat{\uv} | \vv \right)
= \frac{1}{\sqrt{2 \pi \prod_k \sigma_k^2}}
\exp \left( - \frac{1}{2} \sum_k \frac{ \left( \vv_k - \hat{\uv}_k \right)^2 }{\sigma_k^2} \right) \\
&= \frac{1}{\sqrt{2 \pi \prod_k \sigma_k^2}}
\exp \left( - \frac{f^2 h \sec \theta}{2 N_0} \sum_k \Gm_{k,k} \left( \vv_k - \hat{\uv}_k \right)^2 \right)
\end{split}
\end{equation*}
where we have implicitly defined elements
\begin{equation} \label{equation:Gramian}
\begin{split}
\Gm_{k,k} &= \frac{\left( x_{\mathrm{u},k} - x_{\mathrm{l},k} \right)}{2}
\left( \frac{1}{z_{\mathrm{l},k}^2} - \frac{1}{z_{\mathrm{u},k}^2} \right)
= \frac{\tilde{\mathcal{A}_k}}{f^2 h \sec \theta} .
\end{split}
\end{equation}
The maximum likelihood (ML) decision rule for this classification task can be expressed as
\begin{equation} \label{equation:OptimalClassifier}
\hat{\uv}_{\mathrm{ML}} \left( \vv \right) = \arg \min_{\hat{\uv}} \left\| \vv - \hat{\uv} \right\|_{\Gm}.
\end{equation}
In the representation above, $\| \cdot \|_{\Gm}$ is the norm induced by the generalized inner product 
$\langle \wv_1 | \wv_2 \rangle_{\Gm} = \wv_2^{\mathrm{T}} \Gm \wv_1$, where Gramian matrix $\Gm$ is a positive-definite diagonal matrix whose non-zero entries are given by \eqref{equation:Gramian}.

The key insight revealed through this analysis is that a weighted inner product and its induced norm $\| \cdot \|_{\Gm}$ should be used in the localization process rather than the standard inner product and the Euclidean norm.
The weights of the generalized inner product are dictated by the physics of the camera and can accommodate fine or coarse granularity.
At this point, it is appropriate to assess the potential gains associated with this algorithmic improvement.

\subsection{Preliminary Performance Assessment}
\label{subsection:PreliminaryPerformanceAssessment}

In this section, we assess performance for the model presented above.
We assume the rectangular regions have constant magnitude, i.e., $|a_k| = a$ for all locations.
From \eqref{equation:OptimalClassifier}, we gather that the true location is selected by the ML decision rule whenever
\begin{equation} \label{equation:OptimalClassifierSquared}
\left\| \vv - \uv^* \right\|_{\Gm}^2
\leq \left\| \vv - \hat{\uv} \right\|_{\Gm}^2 \quad \forall \hat{\uv} \neq \uv^* .
\end{equation}
If we look at one alternate location at a time, the condition of \eqref{equation:OptimalClassifierSquared} reduces to $\left\| \vv - \uv^* \right\|_{\Gm}^2 - \left\| \vv - \hat{\uv} \right\|_{\Gm}^2 < 0$, or alternatively,
\begin{equation*}
0 < \left\langle \uv^* - \hat{\uv} | \uv^* \right\rangle_{\Gm}
+ \left\langle \uv^* - \hat{\uv} | \nv \right\rangle_{\Gm} .
\end{equation*}
The first component is a known constant, whereas the second component is a zero-mean Gaussian random variable.
The noise component can be rewritten as $\sum_k (u_k^* - \hat{u}_k) \Gm_{k,k} n_k$ and its variance becomes
\begin{equation*}
\begin{split}
\sum_k (u_k^* - \hat{u}_k)^2 \Gm_{k,k}^2 \sigma_k^2
&= \frac{N_0}{f^2 h \sec \theta} \sum_k (u_k^* - \hat{u}_k)^2 \Gm_{k,k} .
\end{split}
\end{equation*}
We can then express the probability of error given a single alternative as
\begin{equation} \label{equation:WeightedCDF}
1 - \boldsymbol{\Phi} \left(\sqrt{ \frac{f^2 h \sec \theta}{N_0} }
\frac{\left\langle \uv^* - \hat{\uv} | \uv^* \right\rangle_{\Gm}}{\left\| \uv^* - \hat{\uv} \right\|_{\Gm}} \right) .
\end{equation}
In comparison, if a standard inner product is employed as the basis for classification, the probability of error becomes
\begin{equation} \label{equation:UnweightedCDF}
1 - \boldsymbol{\Phi} \left(\sqrt{ \frac{f^2 h \sec \theta}{N_0} }
\frac{\left\langle \uv^* - \hat{\uv} | \uv^* \right\rangle}{\left\| \uv^* - \hat{\uv} \right\|_{\Gm^{-1}}} \right) .
\end{equation}

\subsection{Simulated Performance}
\label{subsection:SimulatedPerformance}

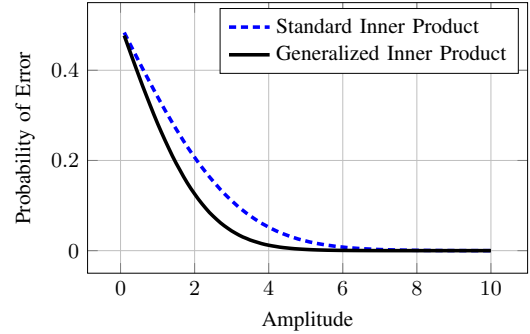
\begin{figure}[tbh]
\centerline{\scalebox{0.9}{\begin{tikzpicture}

\begin{axis}[
font=\small,
scale only axis,
width=6.5cm,
height=4cm,
ymin=-0.05, ymax=0.55,
xlabel={Amplitude},
ylabel={Probability of Error},
xmajorgrids,
ymajorgrids,
zmajorgrids,
legend entries={Standard Inner Product, Generalized Inner Product},
legend style={nodes=right}]

\addplot [color=blue, densely dashed, line width=1.5pt]
coordinates{
(0.1,0.48369224899999996)
(0.2,0.46738853399999997)
(0.3,0.451121999)
(0.4,0.435029111)
(0.5,0.41892366600000003)
(0.6,0.40306986)
(0.7,0.387323885)
(0.8,0.37181707700000005)
(0.9,0.35639884200000005)
(1.0,0.341178297)
(1.1,0.326367099)
(1.2,0.311657946)
(1.3,0.297590271)
(1.4,0.28346820899999997)
(1.5,0.269841581)
(1.6,0.25666072)
(1.7,0.243347368)
(1.8,0.230699375)
(1.9,0.21852212699999998)
(2.0,0.207039904)
(2.1,0.19533240300000002)
(2.2,0.18453090100000002)
(2.3,0.17396621399999998)
(2.4,0.16361800099999999)
(2.5,0.15402238599999998)
(2.6,0.144604816)
(2.7,0.135252768)
(2.8,0.12660429)
(2.9,0.118474915)
(3.0,0.110800934)
(3.1,0.103295814)
(3.2,0.09580240400000001)
(3.3,0.089536995)
(3.4,0.083103089)
(3.5,0.07726971)
(3.6,0.071569998)
(3.7,0.066156987)
(3.8,0.060952578)
(3.9,0.056601927)
(4.0,0.052089777000000004)
(4.1,0.047927578)
(4.2,0.043952837)
(4.3,0.040416331)
(4.4,0.037088181)
(4.5,0.034025365)
(4.6,0.030921145)
(4.7,0.028394721)
(4.8,0.025724132)
(4.9,0.023604878)
(5.0,0.021560304)
(5.1,0.019409362)
(5.2,0.017761053)
(5.3,0.015984895)
(5.4,0.014576553999999999)
(5.5,0.013180003999999999)
(5.6,0.011949558999999998)
(5.7,0.010826533000000001)
(5.8,0.009654179)
(5.9,0.008696139)
(6.0,0.007895807)
(6.1,0.006993984)
(6.2,0.006335911)
(6.3,0.005639543)
(6.4,0.005032347)
(6.5,0.004563772)
(6.6,0.0040019420000000005)
(6.7,0.003628696)
(6.8,0.0032149929999999998)
(6.9,0.002841541)
(7.0,0.002486381)
(7.1,0.002242079)
(7.2,0.00199484)
(7.3,0.001774876)
(7.4,0.00155473)
(7.5,0.001382097)
(7.6,0.0012065539999999999)
(7.7,0.001067815)
(7.8,0.0009407080000000001)
(7.9,0.000834899)
(8.0,0.000736873)
(8.1,0.000637168)
(8.2,0.000562708)
(8.3,0.00048957)
(8.4,0.00043135900000000003)
(8.5,0.000379118)
(8.6,0.000333356)
(8.7,0.000292966)
(8.8,0.00025158)
(8.9,0.000215102)
(9.0,0.000192685)
(9.1,0.000165313)
(9.2,0.000146316)
(9.3,0.00012540700000000001)
(9.4,0.000108441)
(9.5,9.63e-05)
(9.6,8.25e-05)
(9.7,7.5e-05)
(9.8,6.15e-05)
(9.9,5.28e-05)
(10.0,4.5799999999999995e-05)
};

\addplot [color=black, solid, line width=1.5pt]
coordinates{
(0.1,0.476986696)
(0.2,0.454151026)
(0.3,0.43131634700000004)
(0.4,0.40879778299999997)
(0.5,0.38656799299999994)
(0.6,0.364455606)
(0.7,0.343339739)
(0.8,0.32243566)
(0.9,0.3018304)
(1.0,0.282019378)
(1.1,0.263135462)
(1.2,0.24489008899999998)
(1.3,0.22736044)
(1.4,0.21007695699999998)
(1.5,0.194025146)
(1.6,0.179249372)
(1.7,0.164291916)
(1.8,0.150448337)
(1.9,0.137569293)
(2.0,0.1258344)
(2.1,0.114349983)
(2.2,0.103853482)
(2.3,0.094016511)
(2.4,0.084615155)
(2.5,0.076552755)
(2.6,0.068887274)
(2.7,0.061619793)
(2.8,0.055227841)
(2.9,0.049386756)
(3.0,0.043784225)
(3.1,0.039020944)
(3.2,0.034423715)
(3.3,0.030335142000000002)
(3.4,0.027105782000000002)
(3.5,0.023579316)
(3.6,0.020873291000000002)
(3.7,0.01835267)
(3.8,0.015812114)
(3.9,0.013724958999999998)
(4.0,0.012018928)
(4.1,0.010503816000000001)
(4.2,0.009162266)
(4.3,0.007773125)
(4.4,0.006745760999999999)
(4.5,0.005910756999999999)
(4.6,0.005033648)
(4.7,0.004277212)
(4.8,0.003738398)
(4.9,0.003203488)
(5.0,0.0027442390000000003)
(5.1,0.002318499)
(5.2,0.001963419)
(5.3,0.001663559)
(5.4,0.00142662)
(5.5,0.001198857)
(5.6,0.000980612)
(5.7,0.0008748460000000001)
(5.8,0.0007424339999999999)
(5.9,0.0006162359999999999)
(6.0,0.00052203)
(6.1,0.00043247800000000003)
(6.2,0.000369911)
(6.3,0.00029775900000000003)
(6.4,0.000253993)
(6.5,0.00020601400000000002)
(6.6,0.000179127)
(6.7,0.000145241)
(6.8,0.000120235)
(6.9,0.00010921100000000001)
(7.0,8.56e-05)
(7.1,7.04e-05)
(7.2,6.19e-05)
(7.3,4.9e-05)
(7.4,4.2499999999999996e-05)
(7.5,3.32e-05)
(7.6,2.75e-05)
(7.7,2.2100000000000002e-05)
(7.8,1.87e-05)
(7.9,1.6e-05)
(8.0,1.27e-05)
(8.1,1.07e-05)
(8.2,9.58e-06)
(8.3,7.74e-06)
(8.4,6.2e-06)
(8.5,4.86e-06)
(8.6,4.29e-06)
(8.7,3.29e-06)
(8.8,2.76e-06)
(8.9,2.4e-06)
(9.0,1.72e-06)
(9.1,1.91e-06)
(9.2,1.31e-06)
(9.3,1.17e-06)
(9.4,8.5e-07)
(9.5,7.33e-07)
(9.6,6.12e-07)
(9.7,5.21e-07)
(9.8,3.83e-07)
(9.9,4.2799999999999997e-07)
(10.0,3.25e-07)
};

\end{axis}
\end{tikzpicture}}}
\caption{This figure shows the generalized inner product outperforming the standard inner product. SNR increases with amplitude squared.}
\label{figure:SimResultsD8}
\end{figure}

We compare the performance of the standard and generalized inner product in Fig.~\ref{figure:SimResultsD8}.
Performance is defined by the probability of error as described in Section~\ref{subsection:PreliminaryPerformanceAssessment}.
We use the following parameters for observations of 66 squares.

\begin{itemize}
    \item h = 58.3095 cm
    \item $\theta$ = 35.9020$^{\circ}$
    \item Field of View: 39.2962$^{\circ}$ (vertical), 70.5288$^{\circ}$ (horizontal)
    \item Focal Length = 0.0367 cm
    \item Square Side Length = 20 cm
    \item $N_0$ = 0.0018
\end{itemize}

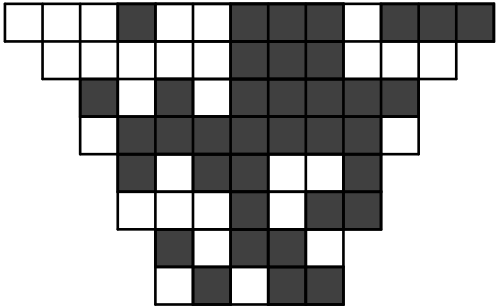
\begin{figure}[tbh]
\centerline{\begin{tikzpicture}[
  font=\small, >=stealth', line width=1.0pt, line cap=round
]

%\foreach \z in {1,2,3,4,5} {
%  \foreach \x in {-2,-1,0,1,2} {
%    \draw (\x, 1) -- (\x, 5);
%    \draw[dashed] (\x, 5) -- (\x, 5.5);
%    \draw (-2, \z) -- (2, \z);
%  }
%}

% fill in squares (L->R) to mimic sample image
% row 1
\draw[fill=darkgray] (-0.75, 0) rectangle (-0.25, 0.5);
\draw[fill=darkgray] (0.25, 0) rectangle (0.75, 0.5);
\draw[fill=darkgray] (0.75, 0) rectangle (1.25, 0.5);
% row 2
\draw[fill=darkgray] (-1.25, 0.5) rectangle (-0.75, 1);
\draw[fill=darkgray] (-0.25, 0.5) rectangle (0.25, 1);
\draw[fill=darkgray] (0.25, 0.5) rectangle (0.75, 1);
% row 3
\draw[fill=darkgray] (-0.25, 1) rectangle (0.25, 1.5);
\draw[fill=darkgray] (0.75, 1) rectangle (1.25, 1.5);
\draw[fill=darkgray] (1.25, 1) rectangle (1.75, 1.5);
% row 4
\draw[fill=darkgray] (-1.75, 1.5) rectangle (-1.25, 2);
\draw[fill=darkgray] (-0.75, 1.5) rectangle (-0.25, 2);
\draw[fill=darkgray] (-0.25, 1.5) rectangle (0.25, 2);
\draw[fill=darkgray] (1.25, 1.5) rectangle (1.75, 2);
% row 5
\draw[fill=darkgray] (-1.75, 2) rectangle (1.75, 2.5);
% row 6
\draw[fill=darkgray] (-2.25, 2.5) rectangle (-1.75, 3);
\draw[fill=darkgray] (-1.25, 2.5) rectangle (-0.75, 3);
\draw[fill=darkgray] (-0.25, 2.5) rectangle (2.25, 3);
% row 7
\draw[fill=darkgray] (-0.25, 3) rectangle (1.25, 3.5);
% row 8
\draw[fill=darkgray] (-1.75, 3.5) rectangle (-1.25, 4);
\draw[fill=darkgray] (-0.25, 3.5) rectangle (1.25, 4);
\draw[fill=darkgray] (1.75, 3.5) rectangle (3.25, 4);

\foreach \x in {-1.25,-0.75,-0.25,0.25,0.75,1.25} {
  \draw (\x, 0) -- (\x, 4);
}

\foreach \z in {0,0.5,1,1.5,2,2.5,3,3.5,4} {
  \draw (-1.25, \z) -- (1.25, \z);
}

% vertical lines
\draw (-1.75, 1) -- (-1.75, 4);
\draw (1.75, 1) -- (1.75, 4);
\draw (-2.25, 2) -- (-2.25, 4);
\draw (2.25, 2) -- (2.25, 4);
\draw (-2.75, 3) -- (-2.75, 4);
\draw (2.75, 3) -- (2.75, 4);
\draw (-3.25, 3.5) -- (-3.25, 4);
\draw (3.25, 3.5) -- (3.25, 4);

% horizontal lines
\draw (-1.75, 1) -- (1.75, 1);
\draw (-1.75, 1.5) -- (1.75, 1.5);
\draw (-2.25, 2) -- (2.25, 2);
\draw (-2.25, 2.5) -- (2.25, 2.5);
\draw (-2.75, 3) -- (2.75, 3);
\draw (-3.25, 3.5) -- (3.25, 3.5);
\draw (-3.25, 4) -- (3.25, 4);

\end{tikzpicture}}
\caption{This grid offers a sample observation of whole squares acquired by the camera (after rectification).
Note the similarity with the photo in Fig.~\ref{figure:ExampleGridImg}.}
\label{figure:SimRoadPlane}
\end{figure}

We randomly generate $\uv^*$ and $\hat{\uv}$ vectors with amplitudes $\pm a$ and calculate the probabilities of error with \eqref{equation:WeightedCDF} and \eqref{equation:UnweightedCDF}.
We perform 10,000 samples for every magnitude, with $a$ ranging from 0.1 to 10.

\subsection{Experimental Results}

Experimental results are obtained using a painted global map and images captured with a camera configuration similar to the parameters listed in Section~\ref{subsection:SimulatedPerformance}.
Preliminary results are aligned with the findings from our simulations.

\begin{figure}[thb]
\centerline{\includegraphics[scale=0.35]{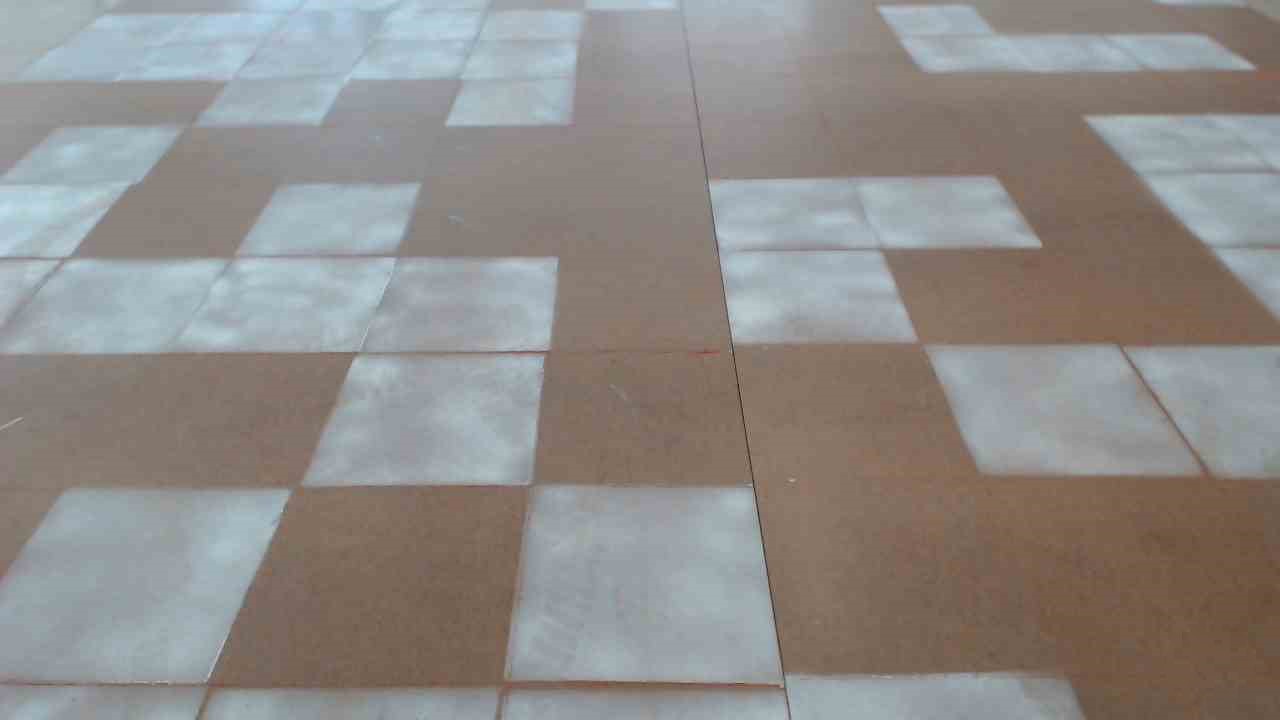}}
\caption{This picture is a sample image of the global map, with its characteristic square sections, taken for our analysis.}
\label{figure:ExampleGridImg}
\end{figure}

\section{Conclusions}

In this article, we discussed the physics of the perspective transformation as it pertains to the acquisition of road images.
We characterized the relationship between the area of a rectangular section of road and its footprint on the focal plane of the camera.
This mapping underlies the phenomenon whereby objects farther from the camera appear smaller than nearer equally-sized objects.
It also affects the signal quality of different sections of the road.
This, in turn, creates an opportunity for an algorithmic improvement for localization tasks that are based on pattern matching.
In particular, we proposed a generalized inner product that takes advantage of this nonuniform signal quality.
This novel generalized inner product is optimal under certain conditions and it outperforms existing methods for localization in noisy conditions.
Our findings are supported by numerical simulations.
The proposed algorithmic framework and its potential improvements are especially significant in harsh environments such as low-light conditions and adverse weather situations, where future autonomous vehicles are poised to operate.

\newpage

\nocite{jo2015precise}
\nocite{rublee2011orb}
\nocite{mur2015orb}
\nocite{engel2014lsd}
\nocite{klein2007parallel}

\bibliographystyle{IEEEbib}
% \bibliography{informedlocalization}

\end{document}